\documentclass[english,pra,aps,twocolumn,floatfix,showpacs,tightenlines,superscriptaddress,amsmath,amssymb]{revtex4}
\usepackage[]{fontenc}
\usepackage[latin1]{inputenc}
\usepackage{verbatim}
\usepackage{graphicx}

\makeatletter

\newcommand{\eq}[1]{Eq.~(\ref{#1})}

\newcommand{\fig}[1]{Fig.~\ref{#1}}

\def\1{1\negthickspace{\rm I}}

\usepackage{babel}
\makeatother
\begin{document}

\title{Long-distance entanglement and quantum teleportation in $XX$ spin chains}

\author{L. Campos Venuti}

\affiliation{ISI Foundation for Scientific Interchange, Villa Gualino, Viale Settimio
Severo 65, I-10133 Torino, Italy}

\author{S. M. Giampaolo}

\affiliation{Dipartimento di Matematica e Informatica, Universit\`{a} degli Studi
di Salerno, Via Ponte don Melillo, I-84084 Fisciano (SA), Italy}

\affiliation{CNR-INFM Coherentia, Napoli, Italy; CNISM Unit\`{a} di Salerno;
and INFN Sezione di Napoli, Gruppo collegato di Salerno, Baronissi
(SA), Italy}

\author{F. Illuminati}

\affiliation{Dipartimento di Matematica e Informatica, Universit\`{a} degli Studi
di Salerno, Via Ponte don Melillo, I-84084 Fisciano (SA), Italy}

\affiliation{CNR-INFM Coherentia, Napoli, Italy; CNISM Unit\`{a} di Salerno;
and INFN Sezione di Napoli, Gruppo collegato di Salerno, Baronissi
(SA), Italy}

\affiliation{ISI Foundation for Scientific Interchange, Villa Gualino, Viale Settimio
Severo 65, I-10133 Torino, Italy}

\author{P. Zanardi}

\affiliation{Department of Physics and Astronomy, University of Southern California
Los Angeles, California 90089-0484, USA}

\affiliation{ISI Foundation for Scientific Interchange, Villa Gualino, Viale Settimio
Severo 65, I-10133 Torino, Italy}


\begin{abstract}
Isotropic $XX$ models of one-dimensional spin-$1/2$ chains are investigated
with the aim to elucidate the formal structure and the physical properties
that allow these systems to act as channels for long-distance, high-fidelity
quantum teleportation. We introduce two types of models: I) open,
dimerized $XX$ chains, and II) open $XX$ chains with small end bonds.
For both models we obtain the exact expressions for the end-to-end
correlations and the scaling of the energy gap with the length of
the chain. We determine the end-to-end concurrence and show that model
I) supports true long-distance entanglement at zero temperature, while
model II) supports \textit{{}``quasi long-distance''} entanglement
that slowly falls off with the size of the chain. Due to the different
scalings of the gaps, respectively exponential for model I) and algebraic
in model II), we demonstrate that the latter allows for efficient
qubit teleportation with high fidelity in sufficiently long chains
even at moderately low temperatures. 
\end{abstract}

\date{August 3, 2007}

\maketitle

\section{Introduction}

The crucial role of entanglement as a fundamental resource for quantum
information tasks that transgress the classical limits is particularly
evident and has been experimentally verified in such protocols like
teleportation \cite{Zeilinger,Boschi}, cryptography and secure key
distribution \cite{Gisin}, and quantum communication \cite{Communication}.
Typically, the {}``most natural way'' to create entanglement between
parties is by means of direct interactions. Since, intuitively, large
amounts of entanglement should be associated to the presence of strong
correlations, low-dimensional systems, as, for instance, spin chains,
offer a natural source of entanglement. However, in most systems with
short-range interactions, the entanglement between a pair of particles
decays rapidly with the distance. For instance, in the Ising model
with transverse field \cite{XX} the two-spin concurrence vanishes
for distances larger than two neighboring sites, while in the Heisenberg
model \cite{Jin}, it is restricted only to single nearest neighbors.
A first exception to this behavior was found by Amico {\em et al}
\cite{Amico}, who established that close to factorization points,
the range of entanglement grows indefinitely. However, in this case,
the entanglement strength between two spins rapidly vanishes as the
range increases. In fact, this behavior appears to be natural, recalling
that monogamy of entanglement \cite{Coffman} implies that if entanglement
is enhanced between two given subsystems, then the entanglement between
any of the two subsystems and a third one must necessarily be suppressed.

From a general quantum informatic perspective, a much desired goal
would be the ability to create a large amount of entanglement between
distant subsystems, simultaneously avoiding direct interactions between
them and single-subsystem addressing. Along this line of thought,
the first requirement can be satisfied by introducing the concept
of localizable entanglement, with the aim of exploiting spin chains
as quantum channels \cite{Verstraete}. The localizable entanglement
measures the average entanglement that can be concentrated on a pair
of distant subsystems by performing optimal local measurements onto
the rest of the system. More recently, it was shown that the ground
state of some spin models with finite correlation length defined 
on one-dimensional chains with open ends can support large values 
of long-distance entanglement between the end points of the chain 
\cite{Bologna}. This approach, i. e.
to look for systems whose ground state can support long-distance entanglement
\textit{prima facie}, without the need of performing operations and
measurements, is clearly very appealing. However, it appears that
models which exhibit true long-distance entanglement are characterized
by energy gaps above the ground state that vanish exponentially as
the length of the chain is increased \cite{Bologna}. Hence, this
interesting phenomenon seems doomed to survive only in the physically
unattainable situation of zero temperature. In the effort to overcome
this problem, it was shown that a kind of long-distance entanglement,
very slowly decreasing with the length of the chain, can be supported
by the ground state of spin models with infinite correlation length defined
on one-dimensional open chains with small end bond interactions,
and that these systems allow for robust finite-temperature teleportation
across finite distances \cite{Bologna2}. This {}``quasi long-distance
entanglement'' between the end points of the chain is associated
to an energy gap that vanishes algebraically with the size of the
system. As a consequence, it is more resilient to thermal excitations,
and can be used to engineer realistic protocols of qubit teleportation,
as has been explicitly demonstrated, by numerical DMRG simulations,
in the case of the Heisenberg ($XXX$) chain with small end interactions
\cite{Bologna2}.

In the present paper we will show that both true long-distance and
quasi long-distance entanglement can be supported by very simple isotropic
$XX$ models of open spin chains. These models present some interesting
advantages over the ones that have been previously studied. At variance
with the Heisenberg case, they are amenable to exact analytical treatment
both in the case of alternating couplings, corresponding to dimerization
of the ground state and true long-distance entanglement, and in the
case of small end bonds, corresponding to a ground state that supports
quasi long-distance entanglement. In this way, one achieves a full
grasp of the physical mechanism responsible for long and quasi-long
distance entanglement, and the possibility to identify unambiguously
the optimal range of parameters for teleportation with maximal fidelity.
Moreover, having in mind possible experimental realizations, for instance
in suitably engineered optical lattices \cite{Duan}, $XX$ chains
are in principle more easily realizable or simulatable than Heisenberg
interactions. The exact solvability of the open-end $XX$ chains is
possible thanks the methods introduced by Lieb and coworkers \cite{Lieb}.
We will first study the fully dimerized $XX$ open spin chain with
alternating couplings, discuss the exact behavior of the end-to-end
concurrence in the ground state at zero temperature, and show analytically
that for this kind of model the energy gap between the ground and
the lowest excited state falls off exponentially with the size of
the system. Therefore, this system cannot be exploited for quantum
teleportation in realistic situations at finite temperature. We then
study the $XX$ open spin chain with small end bonds and determine
analytically the exact expression for the zero temperature end-to-end
concurrence in some physically relevant limits. We determine the exact
scaling of the energy gap with the length of the chain and show that
it dies off algebraically with the size of the system. Finally, we
determine the behavior of the teleportation fidelity as a function
of the temperature for different strengths of the small end bonds,
and show that qubit teleportation with fidelities well above the classical
threshold, and in some cases close to unity, is supported even at
moderately low temperatures.

The paper is organized as follows: In Section II we introduce the
general $XX$ spin chain Hamiltonian with arbitrary site-dependent
couplings, and solve it analytically for the end-to-end two-point
reduced density matrix, correlation, and concurrence (entanglement
of formation). We calculate the fully entangled fraction and establish
the analytical expression for the fidelity of teleportation of an
unknown qubit state between the two end points of the chain. In Section
III we specialize the model to the case of perfectly alternating couplings
and fully dimerized ground state. We determine the single particle
dispersion law, establish the exponential scaling behavior of the
energy gap, and determine the analytical expressions for the ground
state end-to-end concurrence and fidelity. We show that, depending
on the values of the coupling strength, high or even maximal entanglement
and fidelity can be achieved for chains of arbitrary length. In Section
IV we turn to the case of open $XX$ chains with small end bonds,
establishing the single particle dispersion law, the algebraic scaling
of the energy gap, and the analytic expressions for the concurrence
and the fidelity. We then investigate the behavior of the end-to-end
teleportation fidelity as a function of the temperature for different
values of the small end bonds, and conclude that teleportation with
unit fidelity at very low temperature and with high fidelity above
the classical threshold at moderately low temperature are both supported
by the $XX$ channel with small end bonds. Finally, in the conclusions
we summarize our findings and discuss some outlooks on possible future
developments along this line of research.

\section{The general model}

\label{Generalmodel}

As already anticipated in the introduction, we will focus our analysis
on one-dimensional lattices with open ends, described by $XX$ models
with different types of nearest neighbor interactions. Such models
are all special instances of the general $XX$ Hamiltonian 
\begin{equation}
H=\sum_{i=1}^{L-1}J_{i}\left(S_{i}^{x}S_{i+1}^{x}
+ S_{i}^{y}S_{i+1}^{y}\right)\; ,
\label{HamiltonianaGenerica}
\end{equation}
where $J_{i}$ is the interaction strength between nearest neighboring
sites $i$ and $i+1$, $S_{i}^{\alpha}$ are the spin operators defined
at site $i$, and $L$ is the total number of sites (spins) or the
length of the chain. The spectrum of this Hamiltonian can be determined
exactly by a straightforward application of the standard methods introduced
in Ref. \cite{Lieb}. The first step in the procedure is to perform
a Jordan-Wigner transformation \cite{JordanWigner}, 
\begin{eqnarray}
& & S_{i}^{+}=c_{i}^{\dagger}e^{i\pi\sum_{j=1}^{i-1}c_{j}^{\dagger}c_{j}}\;,\qquad S_{i}^{-}
=e^{-i\pi\sum_{j=1}^{i-1}c_{j}^{\dagger}c_{j}}c_{i}\;,\nonumber \\
& & S_{i}^{z}=c_{i}^{\dagger}c_{i}-\frac{\1}{2}\;,
\end{eqnarray}
where $S_{j}^{\pm}=S_{j}^{x}\pm iS_{j}^{y}$. As a result, the Hamiltonian
(\ref{HamiltonianaGenerica}) is mapped in the free fermion Hamiltonian
\begin{equation}
H=\frac{1}{2}\sum_{i=1}^{L-1}J_{i}\left(c_{i}^{\dagger}c_{i+1}+
c_{i+1}^{\dagger}c_{i}\right)=\mathbf{c}^{\dagger}M\mathbf{c}\;,
\label{Trasformata}
\end{equation}
where $\mathbf{c}^{\dagger}=\left(c_{1}^{\dagger},\ldots,c_{L}^{\dagger}\right)$
($\mathbf{c}$) is the vector of the $L$ creation (annihilation)
operators, one for each site of the lattice, and the adjacency matrix
$M$ reads 
\begin{equation}
M=\frac{1}{2}\left|\begin{array}{cccccc}
0 & J_{1} & 0 & \cdots &  & 0\\
J_{1} & 0 & J_{2}\\
0 & J_{2} & 0 &  &  & \vdots\\
\vdots &  &  & \ddots & J_{L-2} & 0\\
 &  &  & J_{L-2} & 0 & J_{L-1}\\
0 &  & \cdots & 0 & J_{L-1} & 0\end{array}\right|\;.
\label{Mgenerica}
\end{equation}
We will investigate two particular realizations of Hamiltonian (\ref{HamiltonianaGenerica})
that, as we will show, in the ground state allow for high or even
maximal entanglement between the end spins of the chain, and thus
naturally provide a channel with high or even unit fidelity for qubit
teleportation. To evaluate the teleportation fidelity, one needs to
determine the spin-spin concurrence (entanglement of formation) between
the end points of the chain. This quantity can be computed exactly
for any two-qubit state (pure or mixed), thanks to the celebrated
formula of Wootters \cite{Wootters}, and the task is left to obtain
its explicit expression in the reduced state of the two end-point
spins. To this purpose, we need to calculate explicitly all the possible
forms of two-point correlations in the ground state. The Hamiltonian
(\ref{HamiltonianaGenerica}) is symmetric under rotations of the
spins around the $z$-axis, so that the only nonvanishing correlations
are $\langle S_{i}^{x}S_{j}^{x}\rangle=\langle S_{i}^{y}S_{j}^{y}\rangle$,
$\langle S_{i}^{z}S_{j}^{z}\rangle$ and $\langle S_{i}^{z}\rangle$.
In the absence of external magnetic fields, $\pi$-rotations around
the $x$ and $y$ axes are symmetries of the model, which additionally
implies $\langle S_{i}^{z}\rangle=0$ at every site. Thanks to the
aforementioned symmetries, the two-point reduced density matrix $\rho_{i,j}$,
obtained by tracing the full density matrix of the system over all
sites except the pair $\{ i,j\}$, has the form 
\begin{equation}
\rho_{i,j}=\frac{\1}{4}+\langle S_{i}^{x}S_{j}^{x}\rangle\left(\sigma^{x}\otimes\sigma^{x}
+ \sigma^{y}\otimes\sigma^{y}\right)+\langle S_{i}^{z}S_{j}^{z}\rangle\sigma^{z}\otimes\sigma^{z},
\label{ReducedDensity}
\end{equation}
where $\sigma^{x,y,z}$ are the Pauli matrices and $\langle\cdot\rangle$
is the ground state average at temperature $T=0$, or the thermal
one at finite temperature $\beta=(k_{B}T)^{-1}$ with respect to the
Gibbs state $\rho=e^{-\beta H}Z^{-1}$. We are interested in the case
in which $i$ and $j$ are the two end points of the chain. In this
instance, we have 
\begin{eqnarray}
& & S_{1}^{+}S_{L}^{-}+S_{1}^{-}S_{L}^{+}=-e^{i\pi\mathcal{N}}\left(c_{1}^{\dagger}c_{L}+
c_{L}^{\dagger}c_{1}\right)\;,\nonumber \\
& & S_{1}^{z}S_{L}^{z}=\left(c_{1}^{\dagger}c_{1}-\frac{1}{2}\right)\left(c_{L}^{\dagger}c_{L}
-\frac{1}{2}\right)\;,
\label{Correlazionigeneriche}
\end{eqnarray}
where $\mathcal{N}=\sum_{i=1}^{L}c_{i}^{\dagger}c_{i}$ is the total
number operator. Using Wick's theorem and taking into account that
$\langle c_{i}^{\dagger}c_{i}\rangle=1/2$, we obtain 
\begin{eqnarray}
& & \langle S_{1}^{+}S_{L}^{-}+S_{1}^{-}S_{L}^{+}\rangle=
-e^{i\pi L/2}\left(\langle c_{1}^{\dagger}c_{L}\rangle
+ \langle c_{L}^{\dagger}c_{1}\rangle\right)\;,\nonumber \\
&  & \langle S_{1}^{z}S_{L}^{z}\rangle=-\langle c_{1}^{\dagger}c_{L}\rangle\langle 
c_{L}^{\dagger}c_{1}\rangle.\end{eqnarray}
Setting $x\equiv\langle c_{1}^{\dagger}c_{L}\rangle$ we see that
the end-to-end reduced density matrix depends uniquely on this parameter.

All the physical informations about model (\ref{Trasformata}) can
now be obtained by diagonalizing the one-body matrix $M$. Let $\xi_{k}$
be the eigenvector with eigenvalue $\Lambda_{k}$, where $k$ is a
quasi-momentum label. Then, passing to new fermionic operators via
the transformation $c_{i}=\sum_{k}\xi_{k}^{\left(i\right)}c_{k}$,
the Hamiltonian takes the form \begin{equation}
H=\sum_{k}\Lambda_{k}c_{k}^{\dagger}c_{k}\;.\end{equation}
The evaluation of the two-point correlation $x$ is then straightforward,
and one obtains 
\begin{eqnarray}
x & = & \sum_{k,q}\xi_{k}^{\left(1\right)}\xi_{q}^{\left(L\right)} \langle 
c_{k}^{\dagger}c_{q} \rangle \nonumber \\
& & \nonumber \\ 
& = & \left\{ \begin{array}{ll}
\sum_{\Lambda_{k}<0}\xi_{k}^{\left(1\right)}\xi_{k}^{\left(L\right)} & \mbox{for }T=0 \\
\rule{0mm}{5mm}\sum_{k}\xi_{k}^{\left(1\right)}\xi_{k}^{\left(L\right)}\frac{1}{1+e^{\beta\Lambda_{k}}} 
& \mbox{for }T>0 \; . \end{array} \right.
\label{correlationgeneral}
\end{eqnarray}
For reduced states of the form (\ref{ReducedDensity}), the end to
end concurrence is easily computed, and we have \cite{Zanardi02}
\begin{equation}
C_{1,L}=2\max\left\{ 0,\, x^{2}+\left|x\right|-\frac{1}{4}\right\} \; .
\label{equazioneconcurrence}
\end{equation}
The above expression of the concurrence is nonvanishing for $|x|>(\sqrt{2}-1)/2\simeq0.207$,
and it reaches the maximum value $C_{1,L}=1$ for $|x|\rightarrow1/2$.

It is natural to expect that the existence of a strong quantum correlation
between the two end spins of the chain can be conveniently exploited
for performing tasks in quantum information, in particular considering
teleportation schemes. In the standard quantum teleportation protocol,
two parties A and B share a maximally entangled state (Bell state). Party
A holds also a third qubit, whose unknown state is to be teleported.
If the two end points of our $XX$ chain share a highly entangled
state, that in some limit may even be asymptotically close to a Bell
state, they can be identified as the two parties, sender and receiver,
for a long-distance, high-fidelity teleportation protocol. The efficiency
of a quantum channel in teleporting an unknown state is quantified
by the fidelity $f$ between the output and the input states, averaged
over all input realizations. The fidelity depends on the actual properties
of the entangled resource $\rho_{1,L}$ (Cfr. \eq{ReducedDensity})
shared by the end spins of the chain. In fact, it has been demonstrated
that the optimal fidelity depends only on the {}``fully entangled
fraction'' $F_{full}$, according to the formula $f=(2F_{full}+1)/3$
\cite{Horodecki}. The fully entangled fraction is defined as the
fidelity between the resource $\rho_{1,L}$ and a maximally entangled
state, maximized over all possible maximally entangled states. For
states of the form \eq{ReducedDensity} it can be easily computed,
and reads $F_{full}=\frac{1}{4}+|x|+x^{2}$ \cite{Badziag}. 
The associated teleportation fidelity is thus 
\begin{equation}
f=\frac{2\left(\frac{1}{4}+|x|+x^{2}\right)+1}{3} \; .
\label{generalfidelity}
\end{equation}
This expression highlights the crucial interplay between entanglement
and efficiency in quantum information protocols. In fact, due to the
high symmetry of states of the form \eq{ReducedDensity}, a nonvanishing
entanglement implies a nonclassical teleportation fidelity exceeding
the classical threshold $2/3$, and viceversa. In the limit $|x|\rightarrow1/2$
of maximally entangled resource, the maximum teleportation fidelity
reaches unity.

\section{Long-distance entanglement}

\label{LDE}

In analogy with previous work on the dimerized Heisenberg model \cite{Bologna},
we consider first the open end $XX$ chain with bonds of alternating
strengths $(1-\delta)$ (weak bond), and $(1+\delta)$ (strong bond),
with $0\le\delta\le1$. \begin{equation}
H = J \sum_{i=1}^{L-1}(1+(-1)^{i}\delta)\left(S_{i}^{x}S_{i+1}^{x} +
S_{i}^{y}S_{i+1}^{y}\right) \; .
\label{LDEHamiltonian}
\end{equation}
Choosing $L$ even and $0\le\delta\le1$ assures that the spins at
the end of the chain interact with a weak bond of strength $(1-\delta)$
with their respective neighbors.

Let us first comment on the general features of the model (\ref{LDEHamiltonian})
in the thermodynamic limit. For $\delta=0$ the model in the fermionic
picture reduces to a simple tight binding with dispersion $\Lambda_{k}=J\cos\left(k\right)$.
The ground state is given by a half filled band and vanishingly small
excitations are present near the Fermi points. Upon the introduction
of a nonvanishing $\delta$ translational invariance by one site is
broken, the Brillouin zone is correspondingly halved, and a gap of
width $\delta$ opens up in the single particle spectrum. Two bands
develop with dispersion 
\begin{equation}
\Lambda_{k,\pm} = \pm J \sqrt{\cos^{2}\left(k\right) +
4\delta^{2}\sin^{2}\left(k\right)} \; .
\label{eq:dimer_bands}
\end{equation}
Upon increasing the value of $\delta$, the growing difference between
the alternating coupling constants allows the creation of dimers between
pairs of strongly interacting spins. The dimers, in turn, are very
weakly interacting with each other. Hence, because the two end spins
of the chain interact very weakly with their nearest neighbors, dimerization
and monogamy of entanglement force the creation of a strong quantum
correlation between the end points, in analogy with what happens in
the Heisenberg case \cite{Bologna}.

The above qualitative picture can be supplemented by a detailed quantitative
analysis both in the thermodynamic limit and at finite length $L$.
The eigenvectors of the adjacency matrix $M$ are given by \cite{Kuznetsova}
\begin{equation}
\xi_{k,\mu_{k}}^{\left(j\right)}=\left\{ \begin{array}{cl}
A_{k}\sin\left(kj\right) & j\,\,\mathrm{even}\;,\\
\mu_{k}A_{k}\sin\left(k\left(L-j+1\right)\right) & j\,\,\mathrm{odd} \; ,
\end{array} \right. \label{LDEeigenvector}
\end{equation}
where $\mu_{k}=\pm1$ is the parity (left-right symmetry) of the
eigenstate $\xi_{k}$, and $A_{k}$ is a normalization constant that
reads 
\begin{equation}
A_{k} = 2 \left[2\left(L+1-\frac{\sin2k\left(L + 
1 \right)}{\sin\left(2k\right)}\right)\right]^{-1/2} \; .
\label{LDEAk}
\end{equation}
The quasimomenta are given by the solution of the following equation
\begin{equation}
\frac{\sin k\left(L+2\right)}{\sin\left(kL\right)}=
-\frac{1+\delta}{1-\delta}\equiv-\frac{1}{a} \; . 
\label{LDEQuasimomenta}
\end{equation}
For each quasimomentum we have two eigenvectors corresponding to
opposite parity. Hence we seek for $L/2$ solutions of (\ref{LDEQuasimomenta}).
When $a<L/\left(L+2\right)$ (roughly $\delta>0$, when $L$ is large)
we have $L/2-1$ real solutions for which the energy dispersion has
the form (\ref{eq:dimer_bands}). The missing solution is complex
(and plays a very important role) and has the form $k_{0}=\pi/2+ip$.
For this mode the equation reads 
\begin{equation}
\cosh\left(2p\right) + \coth\left(Lp\right)\sinh\left(2p\right) =
\frac{1}{a} \; .
\end{equation}
At leading order the solution is given by 
\begin{equation}
e^{2p}=\frac{1}{a}-\left(1-a^{2}\right)a^{L-1} \; .
\end{equation}
The appearance of a complex quasimomentum is crucial for the existence
of long-distance entanglement (LDE) in the ground state and can be
understood as follows. The eigenvectors $\xi_{k_{0},\pm}$ are localized
at the boundaries of the chain. For instance, for site $j$ close
to the left border of the chain ($j\simeq1$), one has 
$\left|\xi_{k_{0},\pm}^{\left(j\right)}\right|\simeq e^{(L-j+1)/\zeta}$,
with a localization length $\zeta$ that, for an asymptotically large
chain, reads 
\begin{equation}
\zeta = 2/\ln\left[\left(1+\delta\right)/\left(1-\delta\right)\right] \; .
\label{localizationlength}
\end{equation}
The energy of this mode, for both value of the parity, is exponentially
small: 
\begin{equation}
\Lambda_{k_{0},\pm}\simeq\pm2\left(1-a\right)e^{-L/\zeta} \; ,
\label{eq:LDE_gap}
\end{equation}
and we thus see that the onset of LDE in the dimerized ground state
of the $XX$ chain with open ends is strictly linked to the existence
of an energy gap above the ground state that vanishes exponentially
fast with the length of the chain. 

\begin{figure}[t]
\includegraphics[width=8cm]{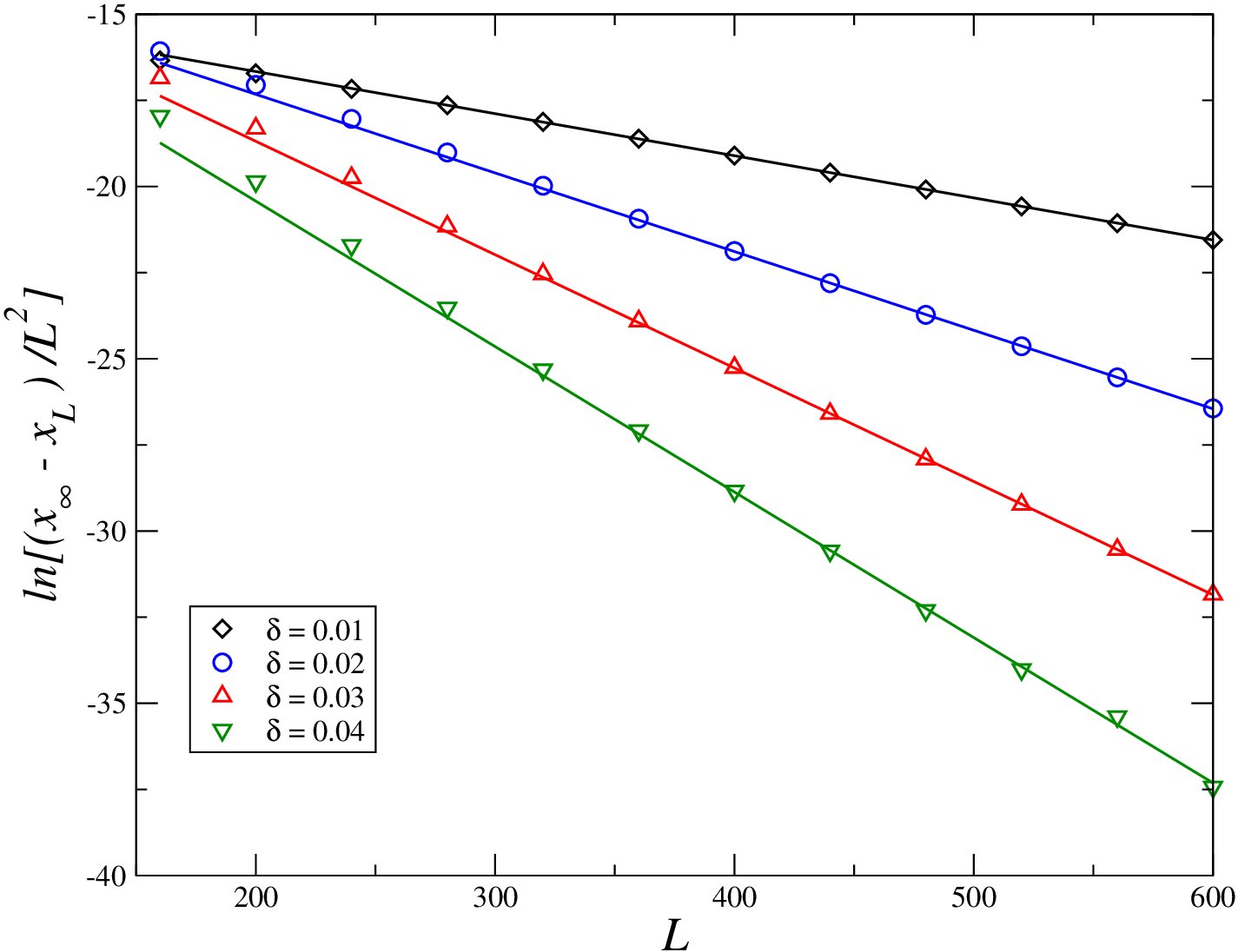} 
\caption{(Color online) Absolute value of the zero-temperature correlation 
$x_{L}=|\langle c_{1}^{\dagger}c_{L}\rangle|$
as a function of the size $L$ of the chain according to the scaling
$x_{L}=x_{\infty}-A_{0}L^{2}\exp(-L/\zeta)$. The different
curves reproduce $\ln[(x_{\infty}-x_{L})/L^{2}]$ as a function of
$L$ for different values of the dimerization parameter $\delta$.
Symbols denote the exact numerical values. The values of $\zeta$ 
are obtained from \eq{localizationlength}.}
\label{figesatta} 
\end{figure}

At zero temperature and zero external field, all negative energy modes
are filled, up to $\Lambda_{k_{0},-}$. The mode $k_{0}$ is responsible
for the appearance of a state localized at the end of the chain in
the many-body ground state. This means that the ground state resemble
$\Psi\simeq\psi_{1,L}\otimes\psi_{\mbox{rest}}$. As we have dimer
order at sites $1$ and $L$, we expect $\psi_{1,L}$ to be highly
entangled. Because factorization will not be exact, $\psi_{1,L}$
will not be exactly a pure state but rather a weakly mixed reduction
$\rho_{1,L}$ that becomes maximally entangled only in the asymptotic
regime of perfect dimerization. In fact, knowing the eigenvectors
of the adjacency matrix, we can evaluate exactly the end-to-end LDE
between the borders of the chain. Using Eq.~(\ref{LDEeigenvector}),
the fermionic correlation function $x$ reads 
\begin{equation}
x = \langle c_{1}^{\dagger}c_{L}\rangle = 
\sum_{k,\,\Lambda_{k}<0}\mu_{k}A_{k}^{2}\left[\sin\left(Lk\right)\right]^{2} \; .
\label{LDEx}
\end{equation}
To evaluate this sum we first isolate the contribution coming from
the complex momentum. The remaining set of terms defines a sum with
alternating signs of a periodic function analytic on the real axis.
The rate of convergence to its asymptotic vanishing value is dictated
by the width of the largest strip, around the imaginary axis, where
the function is analytic. The inverse of this width is precisely given
by the localization length \eq{localizationlength}. Such contribution
is exponentially suppressed, and the final result is 
\begin{eqnarray}
x & = & \left(-1\right)^{L/2}\frac{1}{2}\left(1-a^{2}\right) +
O\left(e^{-L/\zeta}\right) \nonumber \\
& = & \left(-1\right)^{L/2}\frac{2\delta}{\left(1 + 
\delta\right)^{2}} + O\left(e^{-L/\zeta}\right) \; .
\label{endtoendcorrelation}
\end{eqnarray}
We have checked this approximation against exact numerical data and
the results are plotted in Fig.~\ref{figesatta}. It turns out that
the remainder has the form $\left(-1\right)^{L/2}A_{0}L^{2}\exp\left(-L/\zeta\right)
+ \ldots$, where $A_{0}$ is a positive constant. This implies that, in modulus,
the asymptotic value of the end-to-end correlation is in fact approached
from below: In this type of systems, correlations \textit{increase}
as the length $L$ of the chain is increased.

We see that, for any nonvanishing value of the dimerization, the ground
state of the system develops a constant, non-zero correlation between
the end points of arbitrarily large chains, a phenomenon known as
surface order. Using Eq.~(\ref{equazioneconcurrence}), we deduce
that this surface order, present whenever $\delta>0$ (and $\delta<1$)
allows for LDE between the end spins as soon as 
$\delta>\delta_{0} = 
\left(1-\sqrt{2-\sqrt{2}}\right)/\left(1+\sqrt{2-\sqrt{2}}\right)=0.132$,
i.e.~for moderately low values of the dimerization. The corresponding
threshold localization length is $\zeta_{0}=7.479$. The localization
length $\zeta$ rapidly decreases from the threshold value $\zeta_{0}$
with increasing $\delta>\delta_{0}$.
As a consequence, the asymptotic regime for the corresponding end-to-end
concurrence (LDE) is reached already for chains with a size of few
tens of sites. 
Finally, by substituting in \eq{equazioneconcurrence} and in \eq{generalfidelity}
the expression of $x$ \eq{endtoendcorrelation}, we obtain the analytic
expressions of the LDE (end-to-end concurrence) and of the teleportation
fidelity as functions of the dimerization ratio $a=(1-\delta)/(1+\delta)$
in the asymptotic regime of chains of large size $L\gg\zeta_{0}$.
One has: 
\begin{eqnarray}
C_{1,L} & = & 2\max\left\{ 0,\frac{1}{2}-a^{2}+\frac{a^{4}}{4} \right\} \; , \\
& & \nonumber \\ 
f & = & \frac{2}{3}\left[1+\left(\frac{1}{2}-a^{2} +
\frac{a^{4}}{4}\right)\right] \; .
\label{formulefinaliLDE}
\end{eqnarray}
\begin{figure}
\begin{centering} \includegraphics[width=75mm,keepaspectratio]{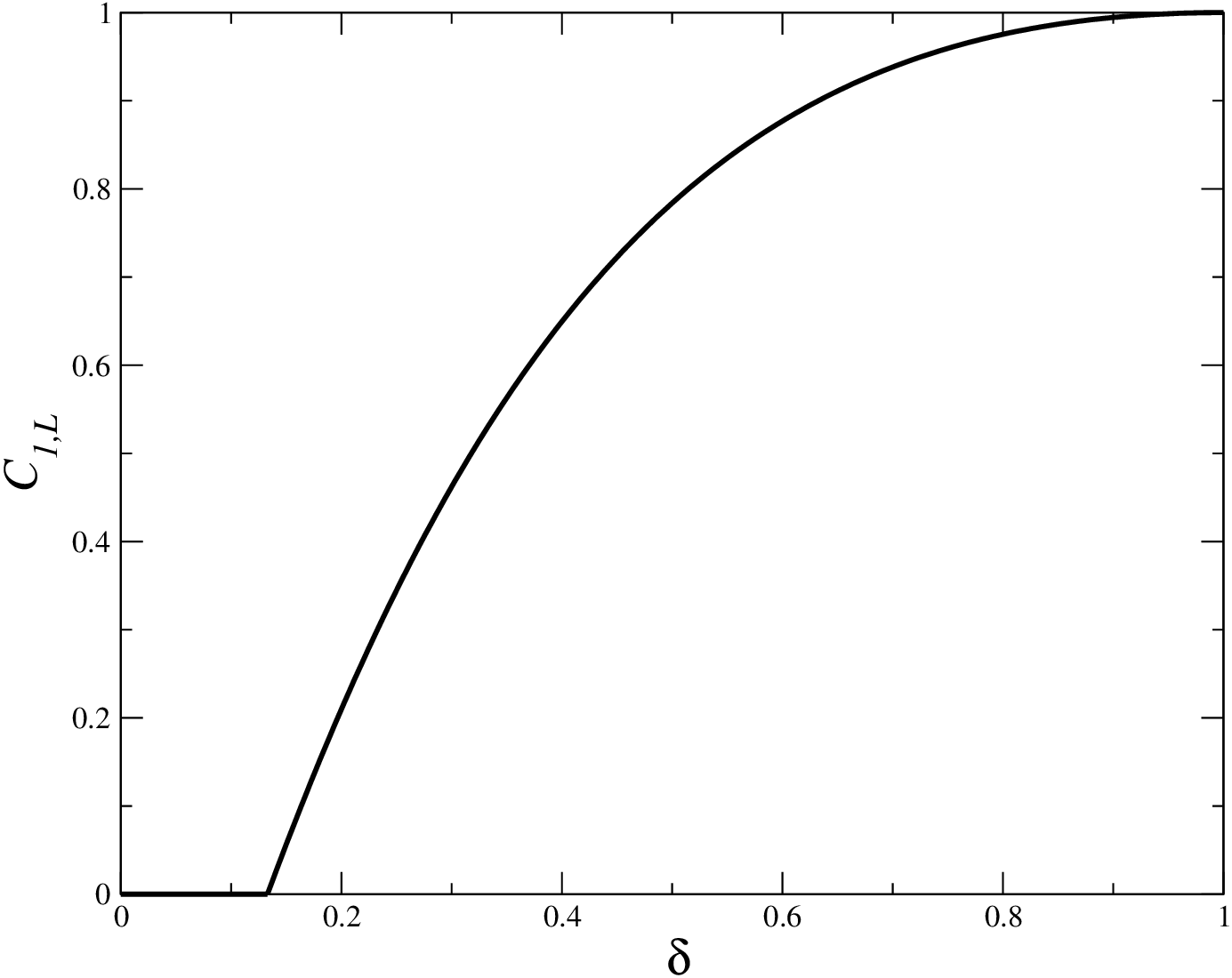}
\par\end{centering}
\caption{End to end concurrence $C_{1,L}$, for long chains of length $L\gg\zeta_{0}$
in the dimer model (\ref{LDEHamiltonian}), as a function of the dimerization
parameter $\delta$. Nonvanishing LDE is achieved and is rapidly growing
as soon as $\delta>\delta_{0}=0.132$.}
\label{fig:concurrence_dimer} 
\end{figure}
In Fig. \ref{fig:concurrence_dimer} we report the behavior of the
LDE between the end points of the chain as a function of the dimerization
parameter $\delta$ for large chains. We see that as soon as the dimerization
is above the threshold value $\delta_{0}$, the LDE grows very rapidly
with $\delta$, reaching saturation in the limiting situation $\delta\rightarrow1$.

Before ending this section we should remark that, although extremely
interesting, the true LDE picture, derived at zero temperature, does
not survive, even qualitatively, at finite temperature. We have seen
that the presence of LDE in the dimerized $XX$ is connected to the
presence of an eigenstate of the adjacency matrix with complex quasimomentum.
From Eq.~(\ref{eq:LDE_gap}) it follows that the energy gap between
the ground state and the lowest excited state vanishes exponentially
as the size of the chain increases. Hence, even for small chains and
very low temperatures, the first excited levels get significantly
populated, contributing to the correlation $x$ with values of opposite
sign with respect to that of the ground state. The total effect is
to lower $x$ down to zero even at very low temperatures. In conclusion,
we have determined that LDE can be supported at zero temperature by
relatively simple, dimerized $XX$ chains with open ends. However,
in close analogy with more complex systems analyzed previously \cite{Bologna},
the zero temperature LDE does not survive as soon as temperature is
switched on.

\section{Quasi long-distance entanglement}

\label{QLDE}

In this section we discuss a second type of model that can support
highly entangled reduced states at the end points of a spin chain.
In all generality, spin models that can allow for strong end-to-end
correlations are characterized by interactions between the end points
and their nearest neighbors that are smaller compared to the interactions
in the bulk of the chain. Otherwise, if the system does not meet this
criterion, the end points would become strongly entangled with their
neighbors, excluding, due to monogamy constraints \cite{Coffman},
the possibility of LDE. We then consider a model of open $XX$ spin
chain formed by $L-2$ spins with uniform coupling strengths, plus
two weakly interacting probes placed at the two end points. Such a
models is described by the following Hamiltonian 
\begin{eqnarray}
H & = & J\sum_{i=2}^{L-2}\left[S_{i}^{x}S_{i+1}^{x}+S_{i}^{y}S_{i+1}^{y} \right.
\label{QLDEHamiltoniana} \\
& &  + \left. \lambda \left(S_{1}^{x}S_{2}^{x}+S_{1}^{y}S_{2}^{y}+
S_{L-1}^{x}S_{L}^{x}+S_{L-1}^{y}S_{L}^{y} \right) \right] \; , \nonumber 
\end{eqnarray}
where $0<\lambda<1$. The presence of the $\lambda$-term in Eq.~(\ref{QLDEHamiltoniana})
can be understood as a kind of generalized boundary condition, reducing
to the standard $XX$ model with uniform couplings and open ends for
$\lambda=0,1$. It is then not surprising that the eigenvalues of
the corresponding adjacency matrix $M$ have the form $\Lambda_{k}=J\cos\left(k\right)$.
The eigenvalue equation determining the quasimomenta for a generic
$\lambda$ reads \cite{Woj05} 
\begin{equation}
\mu_{k}\cot\left(k\right)\left[\cot\left(\frac{L-1}{2}k\right)\right]^{\mu_{k}} =
\, \frac{\lambda^{2}}{2-\lambda^{2}} \; ,
\label{QLDEquasimomenti}
\end{equation}
where $\mu_{k}=\pm1$ is again the parity of the corresponding eigenstate.
The eigenstate $\xi_{k}$ associated to the quasimomentum $k$ has
the following components: 
\begin{eqnarray}
\xi_{k}^{\left(1\right)} & = & \frac{\lambda}{A_{k}}\sin(k) \; , \nonumber \\
\xi_{k}^{\left(i\right)} & = & \frac{1}{A_{k}}\left(\sin((i+1)k)\right. \nonumber \\
& & \left.+(1-\lambda^{2})\sin((i-1)k)\right)\;,\;\;\;1<i<L \; , \nonumber \\
\xi_{k}^{\left(L\right)} & = & \mu\frac{\lambda}{A_{k}}\sin(k) \; , 
\label{eq:QLDE_vector}
\end{eqnarray}
where the normalization $A_{k}$ reads 
\begin{equation}
A_{k}^{2}=(L-1)\left[2(1-\lambda^{2})\cos^{2}(k) + 
\frac{\lambda^{4}}{2}\right]+2\lambda^{2}-\lambda^{4} \, .
\label{QLDEomega}
\end{equation}
Eq.~(\ref{QLDEquasimomenti}) admits $L$ distinct real solutions
in the interval $\left(0,\pi\right)$ for $\lambda\neq0$. Using Eq.~(\ref{eq:QLDE_vector})
the zero temperature end-to-end correlation $x$ reads 
\begin{equation}
x=\sum_{\pi/2<k<\pi}\mu_{k}\frac{\lambda^{2}}{A_{k}^{2}}\sin^{2}k \; .
\label{QLDEx}
\end{equation}
The sum is limited to the $L/2$ quasimomenta that lie in the half-interval
$\pi/2<k<\pi$. This condition takes into account the fact that, at
exactly zero temperature, all levels below the Fermi energy are occupied,
while all the others are empty.
\begin{figure}[t]
\begin{centering} \includegraphics[width=75mm,keepaspectratio]{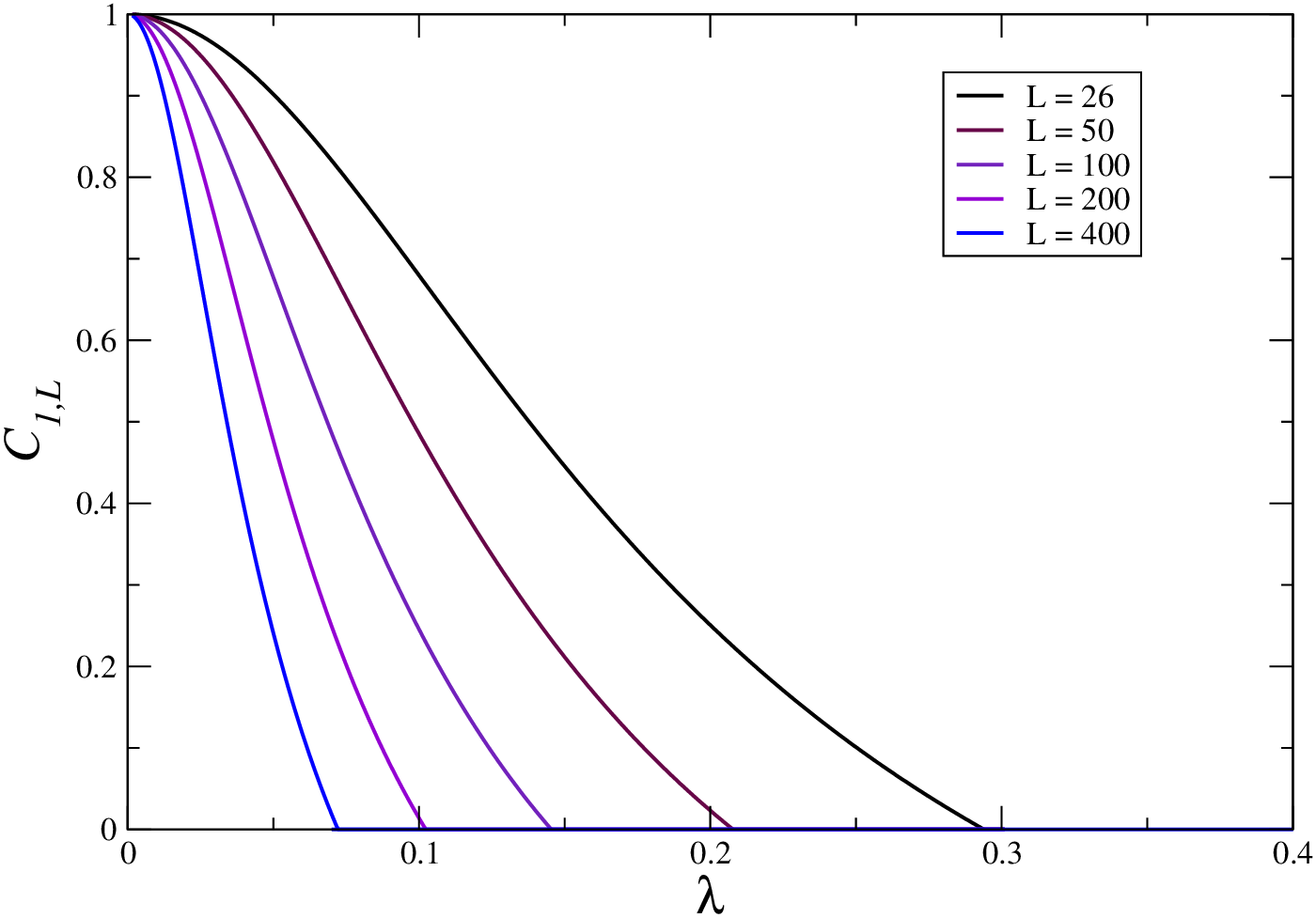} 
\par\end{centering}
\caption{(Color online) End-to-end concurrence $C_{1,L}$ at zero temperature
as a function of $\lambda$ for different lengths $L$ of the chain.
Curves from right to left are, respectively, for $L=26$; $L=50$;
$L=100$; $L=200$; and $L=400$.}
\label{fig:QLDE_conc} 
\end{figure}
The value of the end-to-end concurrence $C_{1,L}$ is readily obtained
by inserting expression (\ref{QLDEx}) for the end-to-end correlation
$x$ in Eq.~(\ref{equazioneconcurrence}). In Fig. \ref{fig:QLDE_conc}
we show the behavior of the end-to-end entanglement as a function
of $\lambda$, for different lengths of the chain. At fixed length
$L$ of the chain, $C_{1,L}\rightarrow1$ as $\lambda\rightarrow0$.
At a fixed value of $\lambda$, the end-to-end concurrence slowly
dies off as $L\rightarrow\infty$. The $XX$ spin chain with small
end bonds is thus characterized by what we may name \textit{{}``Quasi
Long-Distance Entanglement} (QLDE) because, at variance with the dimerized
case considered in the previous section, the end-to-end entanglement,
although able to reach asymptotically maximal values in the limit
of vanishing end-couplings, slowly decreases as the size of the chain
is increased. These considerations can be made quantitative by investigating
in detail the exact expressions for $x$ that are obtained from Eq.~(\ref{QLDEx})
in the two limiting cases $\lambda\rightarrow0,1$.

In the limit of vanishing small end bonds ($\lambda\rightarrow0)$,
the quasimomenta assume the expressions 
\begin{eqnarray}
k_{n} & \simeq & \frac{n\pi}{L-1} - 
\lambda^{2}\frac{\tan\left(\frac{n\pi}{L-1}\right)}{L-1},\quad n=1,2,\dots,L-2 \, ,
\nonumber \\
& & \nonumber \\
k_{\pm} & \simeq & \frac{\pi}{2}\pm\frac{\lambda^{2}}{2} \; .
\label{eq:QLDE_momenta}
\end{eqnarray}
Isolating the dominating contributions in the sum (\ref{QLDEx}),
and keeping terms up to second order in $\lambda$, in the limiting
case $\lambda\rightarrow0$, one obtains 
\begin{equation}
x_{\lambda\rightarrow0}=\left(-1\right)^{L/2}\left[\frac{1}{2} -
\lambda^{2}L\overbrace{\left(\frac{1}{4}+\frac{2\mathcal{C}}{\pi^{2}}\right)}^{c} +
O\left(\lambda^{4}\right)\right] \, , 
\label{eq:QLDE0}
\end{equation}
where $\mathcal{C}$ is Catalan's constant $\mathcal{C}=0.915$,
and the constant in brackets is $c=0.435$. The above result provides
the quantitative content of the qualitative picture sketched above:
At fixed size $L$ of the chain, one can always choose $\lambda$
small enough so that $\left|x\right|\rightarrow1/2$ and the end-to-end
concurrence approaches unity. Depending on the length of the chain,
the condition to be satisfied to achieve large values of the end-to-end
concurrence is $\lambda\ll1/\sqrt{L}$.

\begin{figure}[t]
\begin{centering} \includegraphics[width=75mm,keepaspectratio]{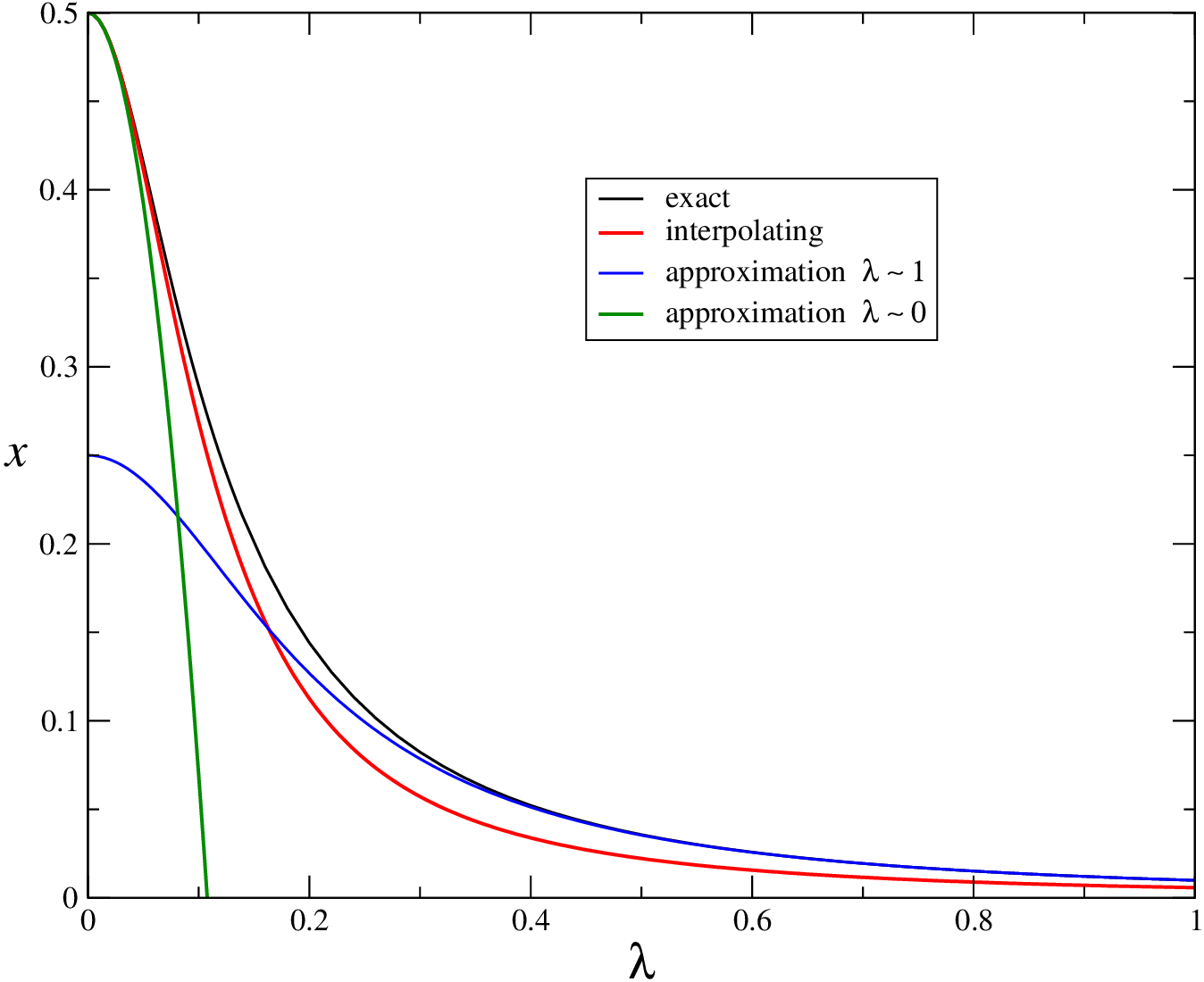} 
\par\end{centering}
\caption{(Color online) Comparison between the exact numerical evaluation
and the analytic approximations for the end-to-end correlation function
$x=\langle c_{1}^{+}c_{L}\rangle$ for a chain of $L=100$ sites.
The intepolating curve is given by $x_{int}=\left(-1\right)^{L/2}/\left(2+4cL\lambda^{2}\right)$.
It is exact up to $O\left(\lambda^{2}\right)$ when $\lambda\rightarrow0$,
but retains its validity also for higher values of the coupling.}
\label{fig:comparison} 
\end{figure}

For completeness, let us now consider the opposite situation of strong
end bonds $\lambda\rightarrow1$. In this case, all the $L$ quasimomenta
can be approximated by $k_{n}\simeq\pi n/(L+1)$ with $n=1,\ldots,L$.
Exploiting the fact that, at leading order, the alternating Riemann
sum of a function is equal to (one-half of) the value of the function
at the extrema, we obtain 
\begin{equation}
x_{\lambda\rightarrow1}\simeq\left(-1\right)^{L/2}\frac{1}{\lambda^{2}\left(L-3\right)+4} \; .
\label{QLDEXapprossimata1}
\end{equation}
In Fig.~\ref{fig:comparison} we compare the exact expression of
$x$ as a function of $\lambda$, obtained by direct numerical diagonalization,
with the two analytic limiting expressions and an interpolating {\it ansatz} 
for a chain of $L=100$ sites.
We will now show that the phenomenon of QLDE is intimately related
to an algebraic scaling behavior of the energy gap that is radically
different from the one, exponentially decreasing with the size of
the chain, exhibited by the fully dimerized $XX$ chain. This feature
turns out to play a crucial role at finite temperature, allowing QLDE
to be more resilient to thermal fluctuations not only at very low
but even at moderately low temperatures.
\begin{figure}[t]
\begin{centering}\includegraphics[width=78mm,keepaspectratio]{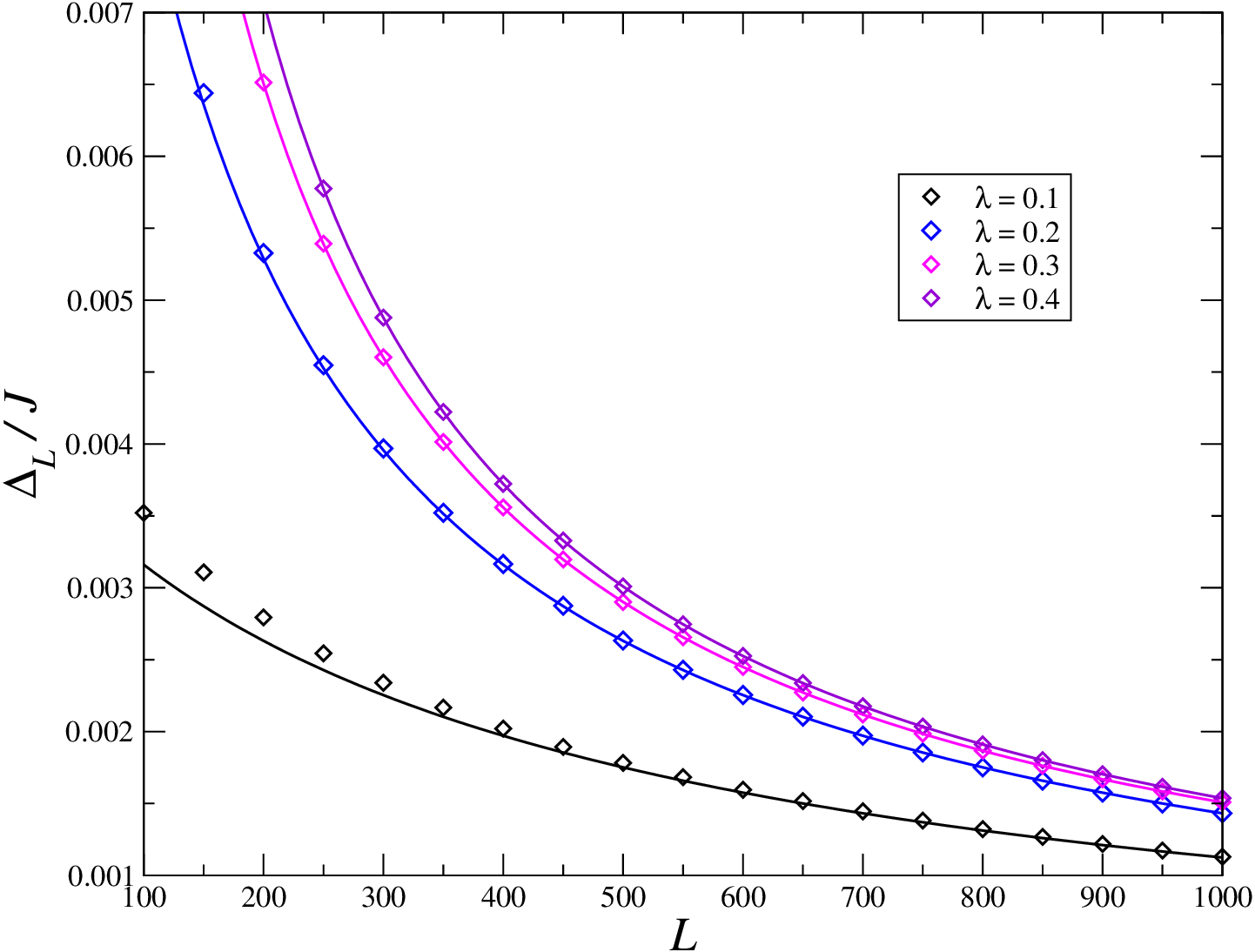} 
\par\end{centering}
\caption{(Color online) Behavior of the energy gap $\Delta_{L}$ as a function
of the length $L$ of the chain for different values of $\lambda$.
Diamonds denote the exact numerical data. Continuous curves display
the approximate expression for the gap (Eq. (\ref{eq:QLDE_gap_approx})
in the text).}
\label{fig:gaps_qlde} 
\end{figure}
As already pointed out, the single-particle dispersion law has the
form $\Lambda_{k}=J\cos k$. The lowest gap is then given by $J\cos\left(k^{\ast}\right)$
where $k^{\ast}$ is the solution of Eq. (\ref{QLDEquasimomenti})
closest to $\pi/2$ with $k^{\ast}<\pi/2$. For sufficiently large
values of $L$, the LHS of Eq. (\ref{QLDEquasimomenti}) has a vertical
asymptote arbitrarily close to $\pi/2$. Expanding the function around
this point we obtain for the desired solution \[
k^{\ast}\simeq\frac{\pi}{2}-\frac{\pi}{2\left(L-1\right)}+
\frac{\pi}{\left(L-1\right)^{2}\lambda^{2}\left(2-\lambda^{2}\right)^{-1} + 
2\left(L-1\right)} \, . \]
Then, at leading order in inverse powers of the length of the chain
$L$, the lowest energy gap reads 
\begin{eqnarray}
& & \frac{\Delta_{L}}{J}\simeq\frac{\pi}{2\left(L-1\right)} -
\frac{\pi}{\left(L-1\right)^{2}\frac{\lambda^{2}}{2 -
\lambda^{2}}+2\left(L-1\right)} \label{eq:QLDE_gap_approx} \\
& & = \frac{\pi}{2}L^{-1}-\left(\frac{\pi}{2} - 
\frac{\pi\left(2-\lambda^{2}\right)}{\lambda^{2}}\right)L^{-2} +
O\left(L^{-3}\right) \, . \nonumber 
\end{eqnarray}
We have compared the asymptotic analytic behaviors for different
$\lambda$ and chains of length up to $L=1000$ sites with the exact
numerical solutions. The results of this comparison are plotted in
Fig.~(\ref{fig:gaps_qlde}), showing excellent agreement of the exact
data with the algebraic scaling (\ref{eq:QLDE_gap_approx}).

The existence of a lowest energy gap slowly falling off algebraically
with the size of the system is to be compared with the exponentially
fast vanishing of the same quantity in the case of the dimerized $XX$
chain. Looking at the structure of the quasimomenta in the two cases,
we immediately realize that the reason for this very different behavior
lies in the absence of complex quasimomenta in the spectrum of the
$XX$ model with small end bonds. On the one hand, this is an undesired
feature, because it is exactly the presence of a complex quasimomentum
that allows for true LDE in the dimerized $XX$ model. On the other
hand this very same feature allows for true LDE and at the same time
is responsible for the exponentially fast vanishing of the energy
gap.

Moving to finite temperature, the exact numerical evaluation of the
fermionic end-to-end correlation function $x$ is obtained using \eq{correlationgeneral}.
After having determined the value of $x$, we can again use Eqs. (\ref{equazioneconcurrence})
and (\ref{generalfidelity}) to evaluate both the end-to-end concurrence
and the teleportation fidelity.
\begin{figure}[t]
\begin{centering}\includegraphics[width=75mm,keepaspectratio]{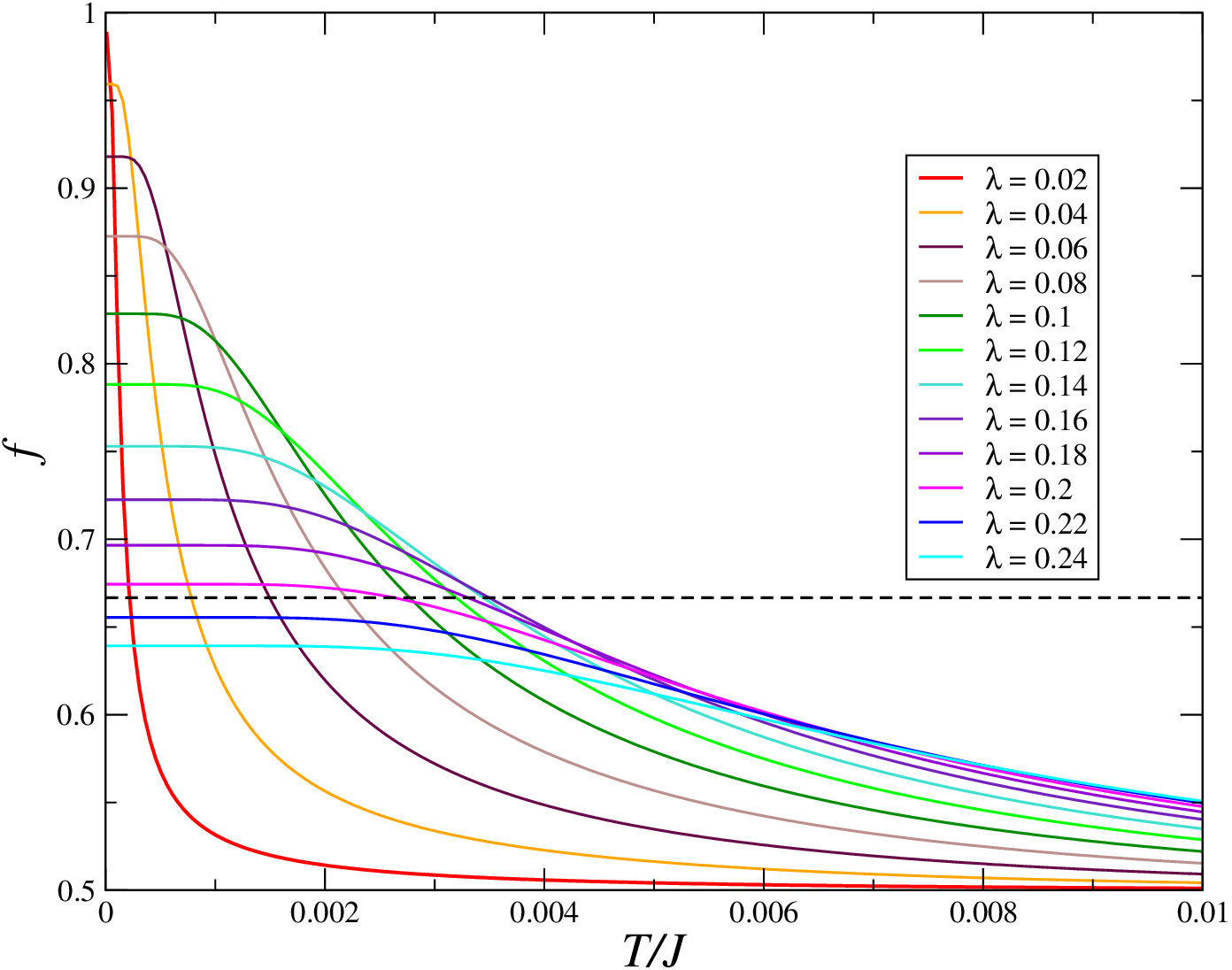} 
\par\end{centering}
\caption{(Color online) Teleportation fidelity for different values of $\lambda$
as a function of the rescaled temperature $T/J$ for a $XX$ chain
with small end bonds, supporting QLDE. Here the length of the chain
is fixed at $L=50$ sites. The horizontal dashed line $f=2/3$ is
the maximum attainable fidelity using a classical teleportation channel.}
\label{fvst} 
\end{figure}
In \fig{fvst} we report the behavior of the teleportation fidelity
$f$ for different values of $\lambda$ as a function of the normalized
temperature $T/J$ for a chain of $L=50$ spins. We see that the higher
the limiting value of the fidelity at vanishing temperature, the faster
$f$ falls off below the classical threshold $f=2/3$. This behavior
is obviously due to the fact that a larger zero-temperature $f$ corresponds
to a smaller energy gap, and hence the system is more sensitive to
the disruptive effect of thermal fluctuations.

We can compare the results reported in \fig{fvst} for an $XX$ chain
with small end bonds with the ones obtained for an Heisenberg antiferromagnetic
chain with two weakly interacting end probes \cite{Bologna2}. The
teleportation efficiency at finite temperature for both models has
been investigated at the same fixed length of the chain $L=50$. From
this comparison, one sees that the behavior of the teleportation fidelity
as a function of the temperature in the two models is qualitatively
similar. This fact is of relevance, because it shows that the QLDE
due to the presence of weakly interacting probes at the and of a uniformly
interacting chain is an effect not restricted to a particular Hamiltonian
model, but can be supported by systems endowed with different bulk
interactions and different symmetries. The only important ingredient
for the realization of a QLDE and a teleportation channel that are
robust against thermal fluctuations is the availability of efficient
control on the interactions at the end points of the channel. Similar
conclusions can be drawn about LDE: In this work we have demonstrated
that existence of true LDE at zero temperature is a phenomenon not
restricted to a particular model and, in fact, does not crucially
depend on the mechanism of dimerization. Preliminary studies indicate
that gapped, anisotropic models in non critical regimes can sustain
LDE even in the absence of a clear pattern of dimerization.

Finally, it is worth noticing that even if the Heisenberg model with
small end bonds analyzed in Ref. \cite{Bologna2} sustains a teleportation
fidelity that remains above the classical threshold for larger values
of the end coupling parameter $\lambda$, the $XX$ channel with small
end bonds assures teleportation with nonclassical fidelities for larger
values of the temperature. Depending on the realistic experimental
settings, if the temperature can be controlled down to sufficiently
low values, it is more convenient to engineer a channel based on the
Heisenberg-like Hamiltonian proposed in Ref. \cite{Bologna2}. Otherwise,
if control on the external temperature cannot be implemented with
high precision down to sufficiently low temperatures, it may then
turn out more convenient to engineer $XX$ chains of the type discussed
in the present paper.

\section{Conclusions and outlook}

In conclusion, we have studied models of open quantum spin chains
endowed with $XX$-like Hamiltonians with nearest-neighbor interactions.
We have discussed two types of models, according to different engineerings
of the spin-spin couplings. In the first case considered, a chain
with bonds of alternating strengths, we have shown how the exact solution
for the ground state implies the existence of long-distance entanglement
between the end spins of the chain, independent of the size of the
system and asymptotically close to unity (maximal entanglement) in
the limit of exact dimerization. This system is therefore perfectly
suited for \textit{bona fide} long-distance quantum teleportation
with ideal fidelity at zero temperature. However, the limiting maximal
values of the fidelity are obtained for an exponentially small energy
gap above the ground state. Therefore, this system is \textit{de facto}
useless for efficient quantum teleportation at finite temperature.
We have then moved on to discuss another class of $XX$ open spin
chains with uniform bulk interactions and small end bonds. In this
case, we have shown that for sufficiently small values of the end
couplings, the ground state of the system supports a quasi long-distance
entanglement between the end spins of the chain, asymptotically close
to unity (maximal entanglement) in the limit of vanishing coupling,
but slowly decreasing as the length of the chain is increased. An
interesting feature of this model is that the lowest energy gap above
the ground state vanishes only algebraically, as the first power of
the inverse of the size of the system. Therefore, in principle, it
can be exploited as a quantum channel for teleportation with nonclassical
fidelity at finite temperature, both very low and moderately low.

Considering future research along these lines of investigation, it
will be interesting to consider practical schemes for the realization
of this kind of spin chains in highly controllable situations, for
instance resorting to ultracold atomic mixtures in optical lattices.
Another interesting open problem worth further study is to assess
the existence and the possible location of a crossover between true
long-distance and \textit{prima facie} quasi long-distance
entanglement behaviors.

\acknowledgments

It is a pleasure to thank Cristian Degli Esposti Boschi and Marco
Roncaglia for many insightful and frutiful discussions. S. M. G. and
F. I. acknowledge financial support from MIUR under PRIN National
Projects 2005, INFN, and CNR-INFM Coherentia. F. I. acknowledges financial
support from the ISI Foundation.


\begin{thebibliography}{99}

\bibitem{Zeilinger} D. Bouwmeester, J.-W. Pan, K. Mattle, M. Eibl,
H. Weinfurter, and A. Zeilinger, Nature \textbf{390}, 575 (1997).

\bibitem{Boschi} D. Boschi, S. Branca, F. De Martini, L. Hardy, and
S. Popescu, Phys. Rev. Lett. \textbf{80}, 1121 (1998).

\bibitem{Gisin} N. Gisin, G. Ribordy, W. Tittel, and H. Zbinden,
Rev. Mod. Phys. \textbf{74}, 145 (2002).

\bibitem{Communication} C. M. Caves and P. D. Drummond, Rev. Mod.
Phys. \textbf{66}, 481 (1994); J. H. Shapiro and O. Hirota (Eds.),
\textit{{}``Quantum Communication, Measurements, and Computing''}
(Rinton Press, Princeton, N.J., U.S.A., 2003).

\bibitem{XX} T. J. Osborne and M. A. Nielsen, Phys. Rev. A \textbf{66},
032110 (2002); A. Osterloh, L. Amico, G. Falci, and R. Fazio, Nature
(London) \textbf{416}, 608 (2002).

\bibitem{Jin} B.-Q. Jin and V. E. Korepin, Phys. Rev. A \textbf{69},
062314 (2004).

\bibitem{Amico} L. Amico, F. Baroni, A. Fubini, D. Patan\`{e}, V.
Tognetti, and P. Verrucchi, Phys. Rev. A \textbf{74}, 022322 (2006).

\bibitem{Coffman} V. Coffman, J. Kundu, and W. K. Wootters, Phys.
Rev. A, \textbf{61}, 052306 (2000); T. J. Osborne and F. Verstraete,
Phys. Rev. Lett. \textbf{96}, 220503 (2006).

\bibitem{Verstraete} F. Verstraete, M. Popp, and J. I. Cirac, Phys.
Rev. Lett. \textbf{92}, 027901 (2004)

\bibitem{Bologna} L. Campos Venuti, C. Degli Esposti Boschi, and
M. Roncaglia, Phys. Rev. Lett. \textbf{96}, 247206 (2006).

\bibitem{Bologna2} L. Campos Venuti, C. Degli Esposti Boschi, and
M. Roncaglia, arXiv:quant-ph/0703202, and Phys. Rev. Lett. (to appear).

\bibitem{Duan} L.-M. Duan, E. Demler, and M. D. Lukin, Phys. Rev.
Lett. \textbf{91}, 090402 (2003).

\bibitem{Lieb} E. Lieb, T. Schultz, and D. Mattis, Ann. Phys. (N.Y.)
\textbf{16}, 407 (1961).

\bibitem{JordanWigner} P. Jordan and E. Wigner, Z. Physik \textbf{47},
631 (1928).

\bibitem{Wootters} W. K. Wootters, Phys. Rev. Lett. \textbf{80},
2245 (1998); S. Hill and W. K. Wootters, Phys. Rev. Lett. \textbf{78},
5022 (1997).

\bibitem{Zanardi02} P.~Zanardi and X.~Wang, J.~Phys.~ A, \textbf{35},
7947 (2002). 

\bibitem{Horodecki} M. Horodecki, P. Horodecki, and R. Horodecki,
Phys. Rev. A. \textbf{60}, 1888 (1999).

\bibitem{Badziag} P.~Badziag, M. Horodecki, P. Horodecki, and R.
Horodecki, Phys. Rev. A. \textbf{62}, 012311 (2000).

\bibitem{Kuznetsova} E. I. Kuznetsova and E. B. Feld'man, JETP \textbf{102},
882 (2006).

\bibitem{Woj05} A. Wojcik, T. Luczak, P. Kurzynski, A. Grudka, T.
Gdala, and M. Bednarska, Phys. Rev. A \textbf{72}, 034303 (2005).

\end{thebibliography}
\end{document}